\documentclass[preprint,5p]{elsarticle} 

\usepackage{lineno}
\usepackage{hyperref}

\usepackage{amsmath}
\usepackage{amsfonts}
\usepackage{amssymb}
\usepackage{physics}
\usepackage{slashed}

\usepackage{graphicx}
\usepackage{float}
\usepackage{subcaption}

\usepackage{widetext}

\usepackage{feynmp-auto}
\newcommand{\half}{\ensuremath{\frac{1}{2}}}
\newcommand{\p}{\ensuremath{\prime}}

\newcommand{\diag}{\ensuremath{\text{diag}}}

\usepackage[utf8]{inputenc}

\modulolinenumbers[5]

\journal{Physics Letters B}

\begin{document}
\begin{fmffile}{feynDiags}

\begin{frontmatter}

\title{Chiral Corrections to Baryon Electromagnetic Form Factors}

\author{Robert J. Perry}
\author{Manuel E. Carrillo-Serrano}
\author{Anthony W. Thomas}
\address{CSSM and ARC Centre of Excellence for Particle Physics at the Tera-scale, Department of Physics, University of Adelaide, Adelaide SA 5005, Australia}

\begin{abstract}
Corrections motivated by chiral symmetry arguments 
have long been known to give important contributions to hadronic observables, 
particularly at low momentum transfer. It is possible to separate these  
approaches into two broad groups; either the corrections are implemented 
at the \emph{parton level}, or at the \emph{hadron level}. We explore the 
results of incorporating pion loop corrections at the hadron level to a 
calculation of electromagnetic form factors in the NJL model. 
These calculations are compared with the result of an earlier implementation 
of pion loops at the parton level using the same NJL model formalism. 
A particular parameter set yields a good description of 
low energy nucleon properties  
within both approaches. However, for the $\Sigma^-$ 
there is a remarkable improvement
when the chiral corrections are implemented at the hadronic level.
\end{abstract}

\end{frontmatter}
%
%

\section{Introduction}\label{sec:1}
The elastic electromagnetic form factors of a hadron are of great interest 
since they are related to the distributions of charge and current 
within the particle. They therefore give vital information about the 
structure of the hadron in question. As a consequence, improved measurements 
of the nucleon electromagnetic form factors have been the subject of many 
ongoing experimental programs. While much good data exists for the 
proton, shorter lifetimes and lack of free targets makes measurements of the  
neutron and other members of the spin-$\half$ baryon octet more difficult. 

After the discovery of QCD, early theoretical 
studies of these form factors were based on quark models, ranging 
from constituent quark models~\cite{Isgur:1987ht,Melde:2007zz} to 
the MIT bag model~\cite{Chodos:1974pn}, with more recent studies employing the  Schwinger-Dyson 
formalism~\cite{Cloet:2013gva}. 
The Nambu--Jona-Lasinio (NJL) model has also been widely 
used~\cite{Ishii:1993np,Schroeder:1999fr} and in this 
paper the quark model component of the 
calculation will be based on earlier 
work~\cite{Cloet:2014rja,Carrillo-Serrano:2016igi}
using the NJL 
model with proper-time regularization~\cite{Ebert:1996vx}
to simulate confinement.

The importance of chiral symmetry, especially for low 
energy hadron properties, was first recognised more 
than 50 years ago in the context of soft pion theorems~\cite{Weinberg:1966kf}, 
while its modern realization is based upon the chiral 
symmetry of QCD itself~\cite{Pagels:1974se}, 
with the pion as a 
pseudo-Goldstone boson, which becomes massless as the 
$u$ and $d$ masses tend to zero. While almost no-one now 
doubts the importance of chiral symmetry and 
particularly the inclusion of the pion as an explicit degree 
of freedom in any quark model calculation, there are still 
important differences between the ways this is implemented.
Here we study the significance of these differences for the 
nucleon, $N$, and $\Sigma$ baryons.

\section{Implementations of Chiral Symmetry}
It is straightforward to understand the correct way to 
implement chiral symmetry. In the chiral limit the pion is 
massless and so a virtual pion can travel infinitely far from 
its source,  {\it provided} the source mass does not change. 
That is, the process $p \rightarrow n \pi^+$ has 
infinite range  
and hence the proton charge radius becomes infinite. 
However, the pion in the process 
$p \rightarrow \Delta^0 \pi^+$ is limited to the range 
$1/\Delta M$, where $\Delta M$ is 
the $\Delta - N$ mass difference in the 
chiral limit (which is similar to the empirical value).

Although simple and completely model independent, 
this argument is often lost in the technicalities of 
building a quark model. For example, it is common 
in spectroscopic studies to introduce pions into the 
quark model Hamiltonian as \cite{Thomas1999}
\begin{equation}
H_{\text{int}}=\frac{g^2}{(4\pi)^2}\frac{1}{3}\sum_{i<j}\vec{\sigma}_i\cdot\vec{\sigma}_j\vec{\tau}_i\cdot\vec{\tau}_j\times\bigg(m_\pi^2\frac{e^{-m_\pi r_{ij}}}{r_{ij}}-4\pi\delta(r_{ij})\bigg),
\end{equation}
where $m_\pi$ is the pion mass. Such models include {\it only} the exchange of 
pions between different quarks.
As shown by Thomas and Krein~\cite{Thomas1999,Thomas2000}, this leads to 
a very large error in the model independent leading 
non-analytic piece of the self-energies of the 
$N$ and $\Delta$ baryons, with the $N/\Delta$ 
ratio being 5 in this model and 1 in chiral perturbation
theory, evaluated as it must be at the {\it hadron level}.

In contrast to this, the use of the {\it parton level} 
approach in previous works has often been motivated by the chiral quark model 
of Georgi and Manohar~\cite{Manohar:1983md}. There one often finds 
pion loops evaluated on individual quarks, rather 
than on the hadron as a whole. It is not hard to see 
that this will also yield the wrong infrared behaviour as 
a function of quark mass. For the proton charge 
radius, for example, the process 
$u \rightarrow d \pi^+$, where $u$ is a valence 
quark in the proton, may leave the three valence
quarks ($udd$) spectator to the pion with either 
spin one half or three halves. As explained above 
these give rise to totally different behaviour for 
the long distance pion cloud.
However, there is no way to keep track 
of those differences if one focusses just on the parton, ignoring
its environment. 

While the LNA behaviour of an observable is a valuable
tool for tracking whether a calculation is 
formally correct, in practice the total pion cloud 
contribution to a particular observable may or may 
not be a bad approximation. Our aim is to investigate 
the practical difference in some interesting examples. 
In particular, we take the NJL model calculations of 
the octet electromagnetic form factors of 
references~\cite{CarrilloSerrano2014,Cloet2014,CarrilloSerrano2016}, 
without pion loops, as the 
bare quark model. We then compare the results for 
low momentum transfer, including charge radii 
and magnetic moments, when the pion corrections 
are performed at the hadron level and at the parton level.

\section{Baryon Form Factors in the NJL Model}\label{sec:2}
The NJL model~\cite{Nambu:1961tp,Nambu:1961fr} 
is a well known constituent quark model. 
The Lagrangian density for the $SU(3)$ flavour NJL Model, 
in its Fierz symmetric form is given as~\cite{CarrilloSerrano2016}
\begin{equation}
\begin{split}
\mathcal{L}=&\overline{\psi}(i\slashed{\partial}-\hat{m})\psi
+\half G_\pi\big[(\overline{\psi}\lambda_i\psi)^2-(\overline{\psi}\gamma_5\lambda_i\psi)^2\big]
\\
&-\half G_\rho\big[(\overline{\psi}\gamma^\mu\lambda_i\psi)^2+(\overline{\psi}\gamma^\mu\gamma_5\lambda_i\psi)^2\big],
\end{split}
\end{equation}
where $\hat{m}=\diag(m_u,m_d,m_s)$ and $\lambda_i$ are the eight generators 
of $SU(3)$ in the Gell-Mann representation, plus $\lambda_0=\sqrt{2/3}$.

Previously, electromagnetic form factors were calculated in the 
NJL Model~\cite{Cloet2014,CarrilloSerrano2016}, where baryons are naturally 
described as quark-diquark bound states~\cite{Cloet2014}. 
The electromagnetic form factors are defined by the matrix elements of the 
electromagnetic current $j^\mu$:
\begin{equation}
\begin{split}
\bra{p^\p,s^\p}j^\mu\ket{p,s}=\overline{u}(p^\p,s^\p)\bigg[&\gamma^\mu F_1^B(Q^2)
\\
&+\frac{i\sigma^{\mu\nu}q_\nu}{2m_B}F_2^B(Q^2)\bigg]u(p,s),
\end{split}
\end{equation}
where $p$ and $s$ designate the momentum and spin states of the baryon, $B$. 
Note that, as is conventional in the literature, $Q^2=-q^2$. In the implementation 
of the NJL Model used here, the quark-photon vertex is dressed by 
including contributions from vector meson dominance. Of particular relevance 
for this paper, that work also showed the effects of pion loops calculated on the 
individual valence quarks.

It is common to use the Sachs parameterization of 
the electromagnetic form factors, which are given as 
linear combinations of $F_1$ and $F_2$;
\begin{align}
G_E^B(Q^2)=&F_1^B(Q^2)-\frac{Q^2}{4m_B^2}F_2^B(Q^2),
\\
G_M^B(Q^2)=&F_1^B(Q^2)+F_2^B(Q^2).
\end{align}
In this parameterization, $G_E^B$ and $G_M^B$ evaluated at $Q^2=0$ are the 
electric charge and magnetic moment of the particle. 
One may also extract the electric charge radius from the slope of $G_E^B$ 
at $Q^2 = 0$.
%
%
The magnetic moments and electric charge radii are well known low energy 
observables, even for the shorter lived spin-$\half$ baryons, and thus 
help to quantify the accuracy of the low $Q^2$ predictions for the 
electromagnetic form factors 
in any model. Importantly, it is at low momentum transfer where 
pions are expected to contribute most and thus differences 
between the implementation of chiral symmetry 
are expected to be most clear.

\section{Chiral Corrections}\label{sec:3}
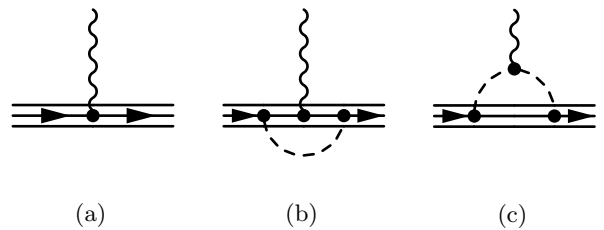
\begin{figure}[b]
\centering
\begin{subfigure}{.3\columnwidth}
\begin{equation*}
\begin{gathered}
\begin{fmfgraph*}(60,80)
\fmfleft{l1}
\fmfright{r1}
\fmftop{t1}
\fmf{fermion}{l1,v1,r1}
\fmffreeze
\fmfi{plain}{vpath (__l1,__v1) shifted (thick*(0,2))}
\fmfi{plain}{vpath (__l1,__v1) shifted (thick*(0,-2))}
\fmfi{plain}{vpath (__v1,__r1) shifted (thick*(0,2))}
\fmfi{plain}{vpath (__v1,__r1) shifted (thick*(0,-2))}
\fmf{photon}{t1,v1}
\fmfdot{v1}
\fmfv{label=$\Gamma^\mu$
,label.angle=45,decor.shape=circle,decor.filled=full,decor.size=2thick,label.dist=10}{v1}
\end{fmfgraph*}
\end{gathered}
\end{equation*}
\vspace{-10mm}
\caption{\label{fig:1.a}}
\end{subfigure}
\begin{subfigure}{.3\columnwidth}
\begin{equation*}
\begin{gathered}
\begin{fmfgraph*}(60,80)
\fmfleft{l1}
\fmfright{r1}
\fmftop{t1}
\fmf{fermion}{l1,v1}
\fmf{plain}{v1,v2}
\fmf{plain}{v2,v3}
\fmf{fermion}{v3,r1}
\fmffreeze
\fmfi{plain}{vpath (__l1,__v1) shifted (thick*(0,2))}
\fmfi{plain}{vpath (__l1,__v1) shifted (thick*(0,-2))}
\fmfi{plain}{vpath (__v3,__r1) shifted (thick*(0,2))}
\fmfi{plain}{vpath (__v3,__r1) shifted (thick*(0,-2))}
\fmfi{plain}{vpath (__v1,__v2) shifted (thick*(0,2))}
\fmfi{plain}{vpath (__v1,__v2) shifted (thick*(0,-2))}
\fmfi{plain}{vpath (__v2,__v3) shifted (thick*(0,2))}
\fmfi{plain}{vpath (__v2,__v3) shifted (thick*(0,-2))}
\fmf{dashes,right=1}{v1,v3}
\fmf{photon}{t1,v2}
\fmfdot{v1,v2,v3}
\fmfv{label=$\Gamma^\mu$
,label.angle=45,decor.shape=circle,decor.filled=full,decor.size=2thick,label.dist=10}{v2}
\end{fmfgraph*}
\end{gathered}
\end{equation*}
\vspace{-10mm}
\caption{\label{fig:1.b}}
\end{subfigure}
\begin{subfigure}{.3\columnwidth}
\begin{equation*}
\begin{gathered}
\begin{fmfgraph}(60,80)
\fmfleft{l1}
\fmfright{r1}
\fmftop{t1}
\fmf{fermion,label=$p$}{l1,v1}
\fmf{plain,label=$p-k$}{v1,v5,v3}
\fmf{fermion,label=$p+q$}{v3,r1}
\fmffreeze
\fmfi{plain}{vpath (__l1,__v1) shifted (thick*(0,2))}
\fmfi{plain}{vpath (__l1,__v1) shifted (thick*(0,-2))}
\fmfi{plain}{vpath (__v3,__r1) shifted (thick*(0,2))}
\fmfi{plain}{vpath (__v3,__r1) shifted (thick*(0,-2))}
\fmfi{plain}{vpath (__v1,__v5) shifted (thick*(0,2))}
\fmfi{plain}{vpath (__v1,__v5) shifted (thick*(0,-2))}
\fmfi{plain}{vpath (__v5,__v3) shifted (thick*(0,2))}
\fmfi{plain}{vpath (__v5,__v3) shifted (thick*(0,-2))}
\fmf{dashes,left=0.4,label=$k$}{v1,v2}
\fmf{dashes,left=0.4,label=$k+q$}{v2,v3}
\fmf{photon,tension=1.6}{t1,v2}
\fmfdot{v1,v2,v3}
\end{fmfgraph}
\end{gathered}
\end{equation*}
\vspace{-10mm}
\caption{\label{fig:1.c}}
\end{subfigure}
\caption{Diagrams which contribute to the calculation of 
electromagnetic form factors. Note that contributions from $\Delta$ intermediate states are not considered in this calculation.
\label{fig:1}}
\end{figure}

For our present purposes the light-front cloudy bag 
model (LFCBM), developed by Miller~\cite{Miller2002} is a 
suitable formalism for calculating the pion cloud 
corrections to baryon form factors.
As in the original cloudy bag 
model~\cite{Theberge1980,Miller1980,Miller:1984em,Thomas1984}, in the 
LFCBM the pion corrections are evaluated at the hadron 
level and therefore ensure the correct LNA behaviour.
Whereas Miller employed a quark model developed by 
Schlumpf~\cite{Schlumpf92}, as noted earlier we use the NJL model. 
Within Miller's approach, using pseudoscalar coupling 
of the pion to the hadron, one obtains three  
Feynman diagrams at one loop order, as shown in Fig.~\ref{fig:1}.

The first diagram (Fig.~\ref{fig:1.a}) is simply the quark model result, while the chiral 
corrections to these form factors are provided by the diagrams shown 
in Figs.~\ref{fig:1.b} and~\ref{fig:1.c}. As the name of the model 
suggests, these equations are evaluated on the light front, 
but importantly, since the form factors $F_1$ and $F_2$ are 
Lorentz invariant scalar functions, it is entirely consistent to take the 
results of the NJL model as input here. 
The results of that work are summarized here as\footnote{A version of 
these equations also exists in~\cite{Miller2002}, but there are 
several small changes in the definitions used here.}~\cite{Matevosyan2005}: 
\begin{equation}
F_{i}^H(Q^2)=Z\big[F_{i,a}^{H}(Q^2)+F_{i,b}^{H}(Q^2)+F_{i,c}^{H}(Q^2)\big] \, ,
\end{equation}
where $i=1,2$, $H$ is the hadron in question, $a$, $b$ and $c$ 
refer to diagrams~\ref{fig:1.a},~\ref{fig:1.b} and~\ref{fig:1.c}, 
respectively, and $Z$ is the wavefunction renormalisation constant, 
defined to ensure that the charge of the proton is unity. 

Evaluation of diagrams given in Figs.~\ref{fig:1.b} and~\ref{fig:1.c} lead to
\begin{widetext}
\begin{align}
\begin{split}
F_{1,b}^H(Q^2)=&\frac{g_{NN\pi}^2}{(4\pi)}\int_0^1dxx\int\frac{d^2L}{(2\pi)^2}\bigg[F_1(Q^2)\bigg(L^2+x^2m_N^2-\frac{1}{4}x^2Q^2\bigg)
-F_2(Q^2)\bigg(\frac{x^2Q^2}{2}\bigg)\bigg]\frac{1}{D(\vec{L}_+^{~2},x)D(\vec{L}_-^{~2},x)},
\end{split}
\\
\begin{split}
F_{2,b}^H(Q^2)=&-\frac{g_{NN\pi}^2}{(4\pi)}\int_0^1dxx\int\frac{d^2L}{(2\pi)^2}
\bigg[F_1(Q^2)\big(2x^2m_N^2\big)
+F_2(Q^2)\bigg(L^2+x^2m_N^2-\frac{1}{4}x^2Q^2\bigg)\bigg]\frac{1}{D(\vec{L}_+^{~2},x)D(\vec{L}_-^{~2},x)},
\end{split}
\end{align}
and
\begin{align}
\begin{split}
F_{1,c}^H=\frac{g_{NN\pi}^2}{(4\pi)}I_\tau F_\pi(Q^2)\int_0^1dxx\int\frac{d^2K}{(2\pi)^2}\bigg[K^2+x^2m_N^2-\frac{1}{4}(1-x)^2Q^2\bigg]\frac{1}{D(\vec{K}_+^{~2},x)D(\vec{K}_-^{~2},x)},
\end{split}
\\
\begin{split}
F_{1,c}^H=\frac{g_{NN\pi}^2}{(4\pi)}I_\tau(2m_N^2)F_\pi(Q^2)\int_0^1dxx^2(1-x)\int\frac{d^2K}{(2\pi)^2}\frac{1}{D(\vec{K}_+^{~2},x)D_N(\vec{K}_-^{~2},x)},
\end{split}
\end{align}
\end{widetext}
where $F_1$ and $F_2$ are given as
\begin{equation}
F_i=
\begin{cases}
2F_{i,a}^n+F_{i,a}^p, \text{ for the proton}
\\
2F_{i,a}^p+F_{i,a}^n, \text{ for the neutron}
\end{cases},
\end{equation}
and $g_{NN\pi}$ is the nucleon-pion coupling constant. In this work 
we take  $Zg_{NN\pi}^2/(4\pi)=13.5$. $D$ is given as
\begin{equation}
D(l_\perp,x)=l_\perp^2+x^2m_N^2+(1-x)m_\pi^2 \, ,
\end{equation}
where $\vec{L}_\pm=\vec{L}_\perp\pm\half x\vec{q}_\perp$. 
The nucleon-pion isospin coupling $I_\tau$ is given as
\begin{equation}
I_\tau=
\begin{cases}
2, \text{ for the proton}
\\
-2, \text{ for the neutron}
\end{cases},
\end{equation}
and $\vec{K}_\pm=\vec{K}_\perp\pm\half(1-x)\vec{q}_\perp$. 

Note that these equations are divergent and require a regularization 
prescription to render them finite. In this work, we choose to use
a $t$-dependent form factor (with $\Lambda$ the regulator mass parameter), 
given as
\begin{equation}
F(\vec{k}_\perp,x)=\exp[-\frac{D(\vec{k}_\perp^{~2},x)}{(1-x)\Lambda^2}] \, ,
\end{equation}
to regulate the formally divergent integrals.
This choice corresponds to the prefered form of regulator in a recent 
study~\cite{McKenney:2015xis} of the origin of the $\bar{d} - \bar{u}$ 
asymmetry in the proton arising from chiral 
effects~\cite{Thomas:1983fh,Melnitchouk:1998rv}.

\subsection{\texorpdfstring{$\Sigma$ Baryons}{Sigma Baryons}}
As explained earlier, we examine not only the chiral corrections 
to nucleon form factors
but also to the $\Sigma$ hyperons. The motivation for this lies in the recent 
work of Carrillo-Serrano {\it et al.}~\cite{CarrilloSerrano2016}, 
which extended the earlier calculation of the nucleon electromagnetic form factors~\cite{Cloet2014}
to the baryon octet. Following the work of de Swart~\cite{deSwart1963}, 
under the assumption of $SU(3)$ flavour symmetry, one has various relationships 
between the couplings of the baryons. The couplings relevant to this work are
\begin{equation}
g_{\Lambda\Sigma\pi}=\frac{2}{\sqrt{3}}(1-\alpha)g_{NN\pi} \, \,; \, \,
g_{\Sigma\Sigma\pi}=2\alpha g_{NN\pi} \, ,
\end{equation}
where we set $\alpha=2/5$. One may then show that the modifications to 
the above equations, in order to evaluate the hyperon form factors, 
are as follows:
\begin{align}
m_N&\rightarrow m_H,
\\
F_i&=4\bigg[\frac{(1-\alpha)^2}{3}F_{i,a}^\Lambda+\alpha^2F_{i,a}^{\Sigma^0}+\alpha^2F_{i,a}^{\Sigma^-}\bigg],
\\
I_\tau&=4\bigg[\frac{(1-\alpha)^2}{3}+\alpha^2\bigg].
\end{align}
Note that we take the $\Sigma^0$ and $\Lambda$ to be mass degenerate in 
the calculation of loop diagrams.
The calculation of baryon electromagnetic form factors in this paper may 
be summarised as a two step process: firstly calculate bare electromagnetic 
form factors in the NJL Model (form factors without the effects of the pion cloud); secondly modify the bare form factors by 
incorporating pion cloud effects in an effective baryon-pion Lagrangian. 

\subsection{Incorporating the Self Energy}
In calculating the chiral corrections, one is effectively including degrees 
of freedom previously absent from the system. These degrees of freedom 
modify bare quantities.
In a self-consistent calculation, the inclusion of 
the pion cloud must also lead to corrections to other observables. 
In particular, the pion cloud also contributes to the baryon self energy, 
which is related to the bare baryon mass $m^{(0)}$ 
via the well-known renormalisation condition
\begin{equation}
m_B=m_B^{(0)}+\Sigma(\slashed{p})\big|_{\slashed{p}=m_B},
\end{equation}
where $m_B$ and $m_B^{(0)}$ are the physical and bare masses of some baryon, 
and $\Sigma(\slashed{p})$ is the self energy. 
As a consequence, the baryon masses in the bare NJL model should no longer 
be the physical masses but rather masses shifted such that  
with inclusion of the self-energy the physical baryon mass is obtained.

\begin{figure}
\centering
\begin{subfigure}{.45\columnwidth}
\begin{equation*}
\begin{gathered}
\begin{fmfgraph*}(80,80)
\fmfleft{l1}
\fmfright{r1}
\fmf{fermion,tension=2}{l1,v1}
\fmf{plain}{v1,v2}
\fmf{fermion,tension=2}{v2,r1}
\fmf{dashes,left=1,tension=0,label=$\pi^\pm$}{v1,v2}
\end{fmfgraph*}
\end{gathered}
\end{equation*}
\vspace{-10mm}
\caption{}\label{fig:2.a}
\end{subfigure}
\begin{subfigure}{.45\columnwidth}
\begin{equation*}
\begin{gathered}
\begin{fmfgraph*}(80,80)
\fmfleft{l1}
\fmfright{r1}
\fmf{fermion,tension=2}{l1,v1}
\fmf{plain}{v1,v2}
\fmf{fermion,tension=2}{v2,r1}
\fmf{dashes,left=1,tension=0,label=$\pi^0$}{v1,v2}
\end{fmfgraph*}
\end{gathered}
\end{equation*}
\vspace{-10mm}
\caption{}\label{fig:2.b}
\end{subfigure}
\caption{Pion contributions to the hadron self energy arise from 
the charged (\ref{fig:2.a}) and neutral (\ref{fig:2.b}) pion 
interactions.}
\label{fig:2}
\end{figure}
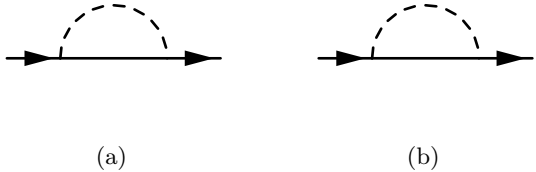

Using the effective field theory discussed above, the self energy 
contribution from the pion to the nucleon must be calculated. 
The relevant diagrams are shown in Fig.~\ref{fig:2}. 
The contribution where 
a form factor $F$ has been used to regularize 
the loop integrals is:
\begin{equation}
\begin{split}
\Sigma(\slashed{p}=m_N)=I_\tau\frac{Zg_{NN\pi}^2}{(4\pi)^2}&\frac{1}{4m_N}\int_0^\infty dt\frac{t|F(-t)|^2}{(t+m_\pi^2)}
\\
&\times\bigg(\frac{t}{m_N^2}-\sqrt{\frac{t^2}{m_N^4}+\frac{4t}{m_N^2}}\bigg).\label{eq:38}
\end{split}
\end{equation}

Numerous studies within the cloudy bag model~\cite{Theberge1982}, Dyson-Schwinger 
equations~\cite{Hecht:2002ej} and lattice motivated 
studies of the $\Delta-N$ mass difference 
as a function of quark mass~\cite{Young:2002cj}, suggest that the self-energy 
contribution from the process $N \rightarrow N \pi$ is of the order 100-150 MeV.
As an illustration, we choose the regulator mass to fix this self-energy at 
130 MeV (so $\Lambda$ = 0.72 GeV).
%

\section{Results}\label{sec:4}
%
\begin{figure*}
\centering
\begin{subfigure}{0.49\textwidth}
\centering
\includegraphics[scale=0.35]{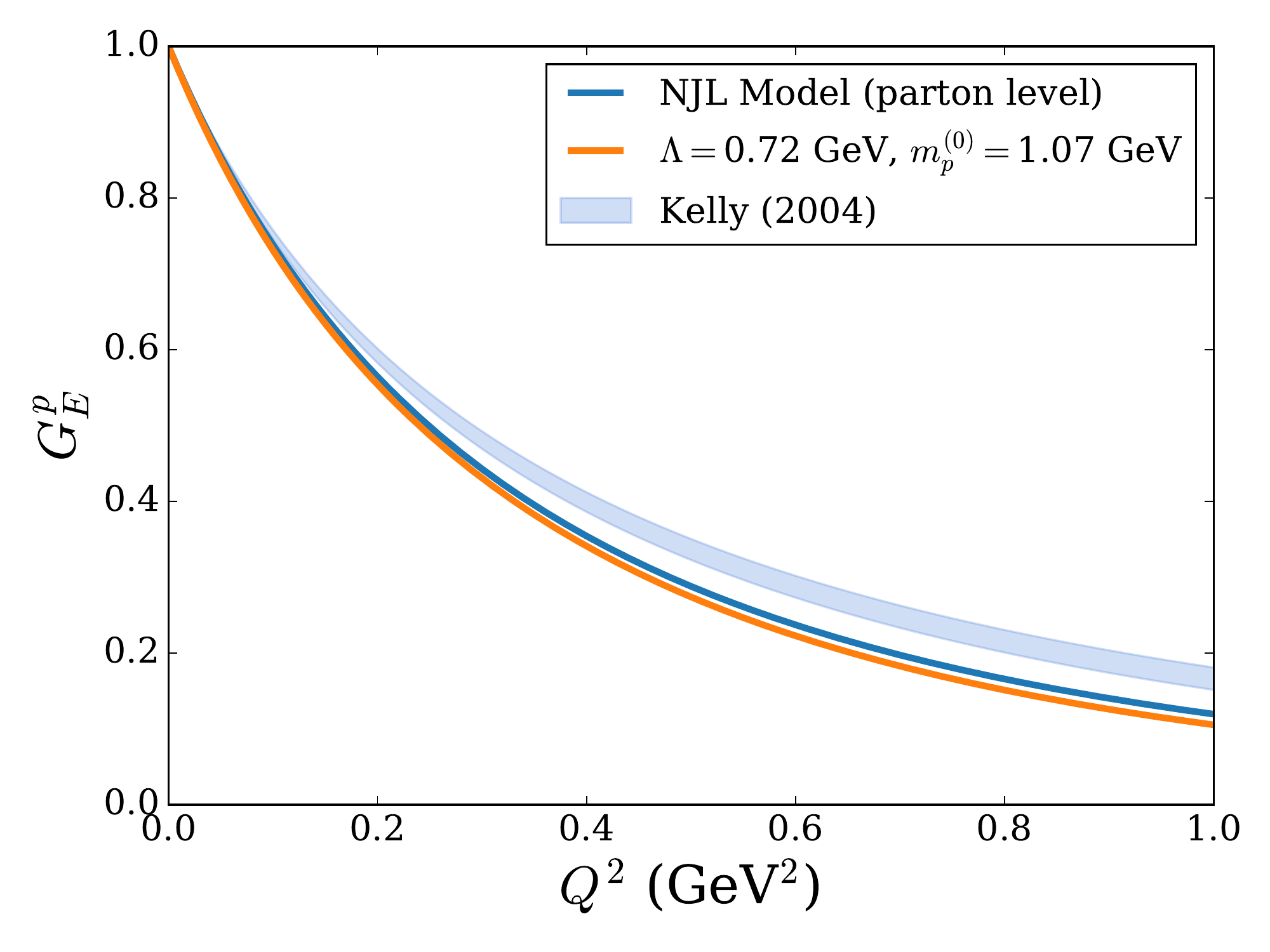}
\end{subfigure}
\begin{subfigure}{0.49\textwidth}
\centering
\includegraphics[scale=0.35]{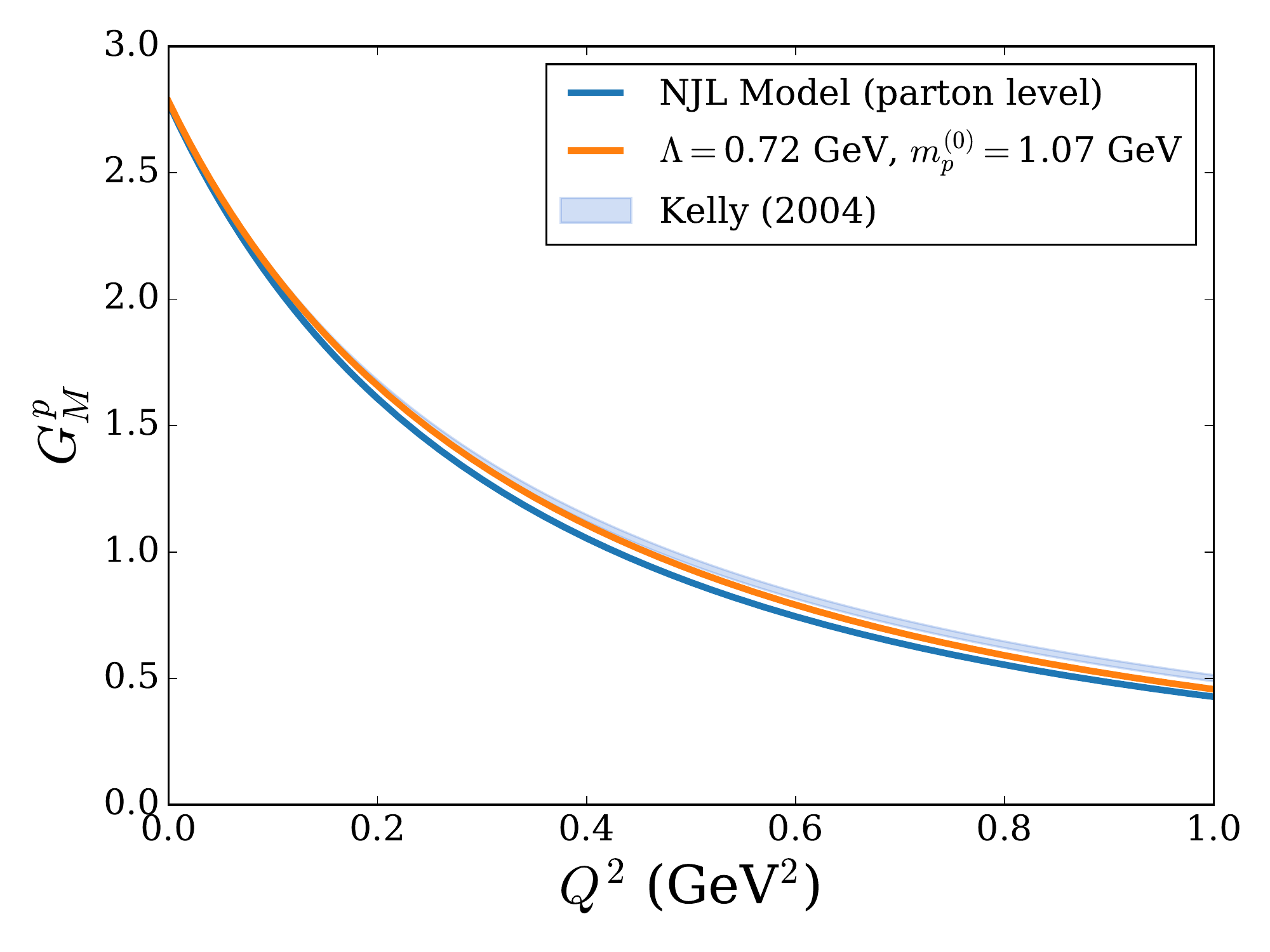}
\end{subfigure}
\begin{subfigure}{0.49\textwidth}
\centering
\includegraphics[scale=0.35]{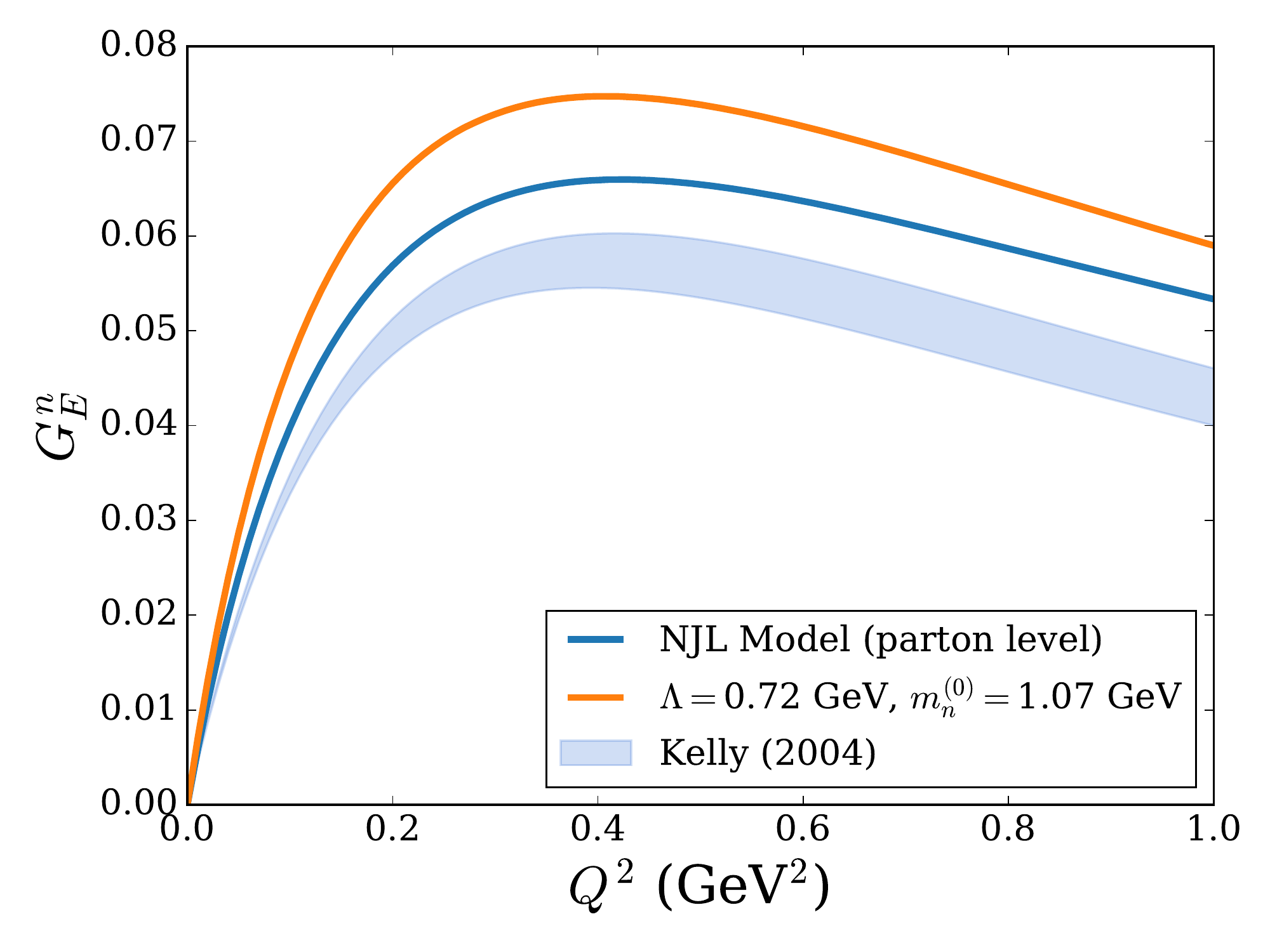}
\end{subfigure}
\begin{subfigure}{0.49\textwidth}
\centering
\includegraphics[scale=0.35]{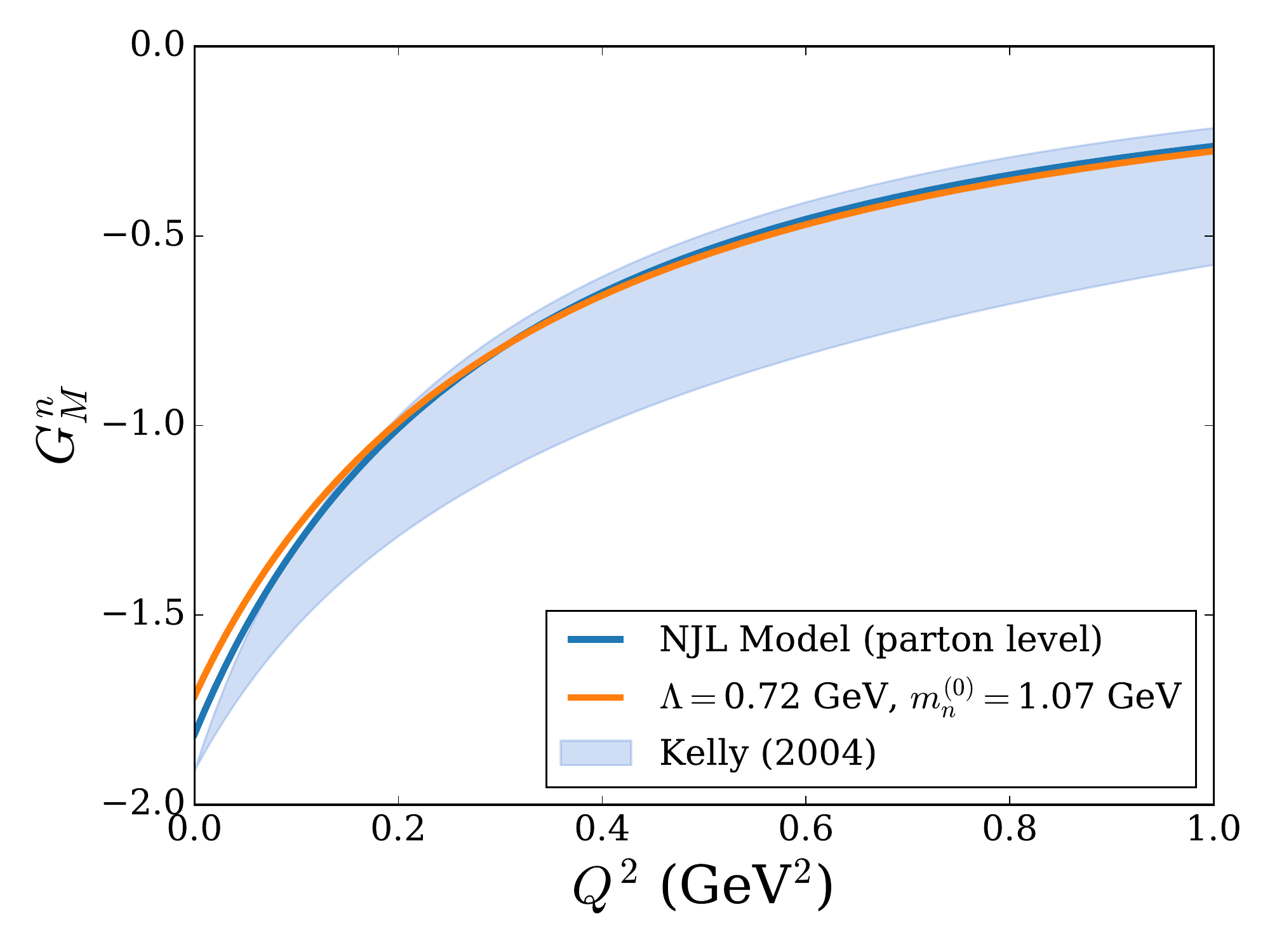}
\end{subfigure}
\vspace{-3mm}
\caption{(Colour online) Electromagnetic form factors for the nucleon. 
The shaded region is obtained from uncertainties in the fitted 
parameters of Kelly's empirical model~\cite{Kelly2004}.}
\label{fig:3}
\end{figure*}
\begin{figure*}
\begin{subfigure}{0.49\textwidth}
\centering
\includegraphics[scale=0.35]{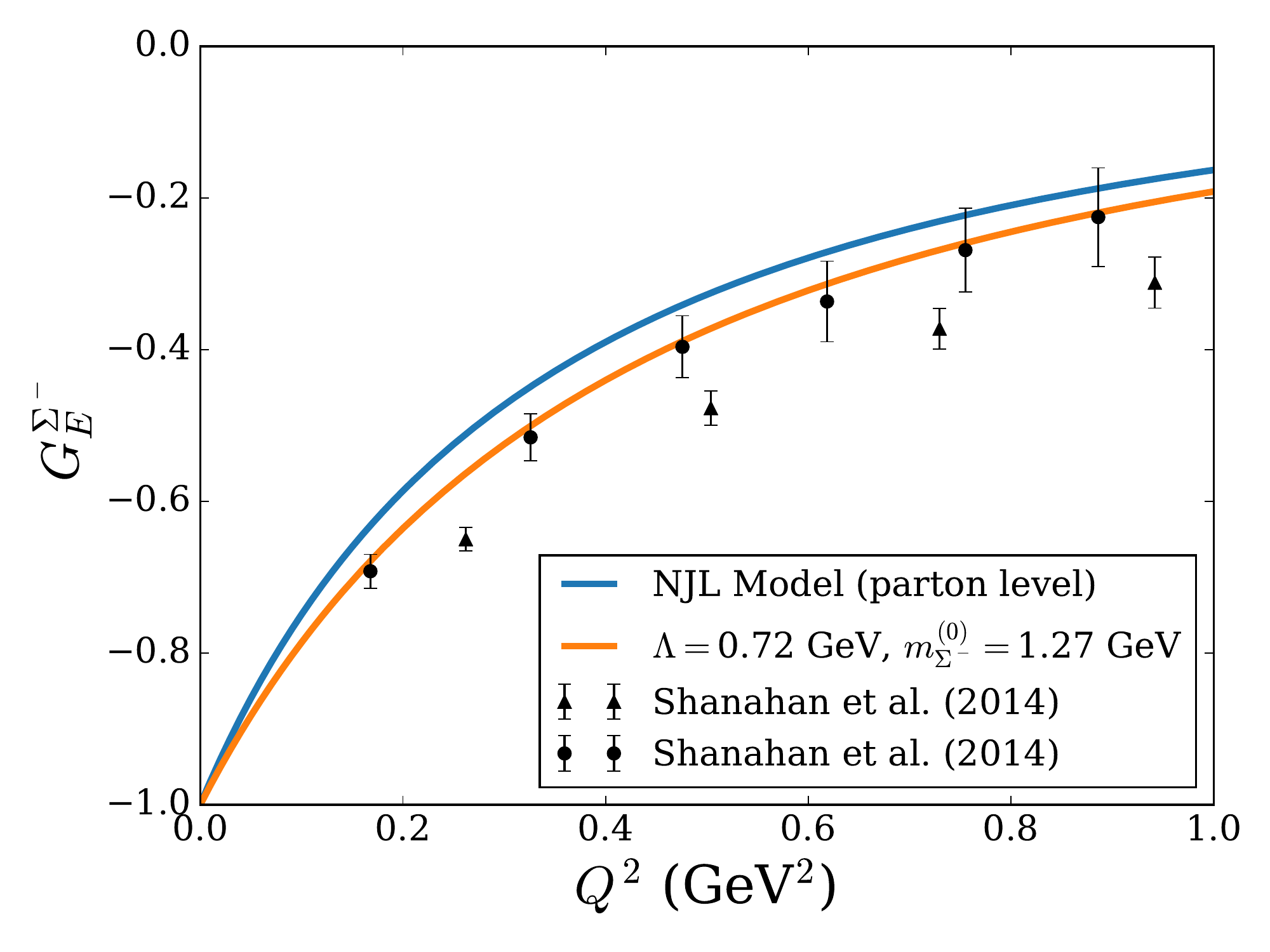}
\end{subfigure}
\begin{subfigure}{0.49\textwidth}
\centering
\includegraphics[scale=0.35]{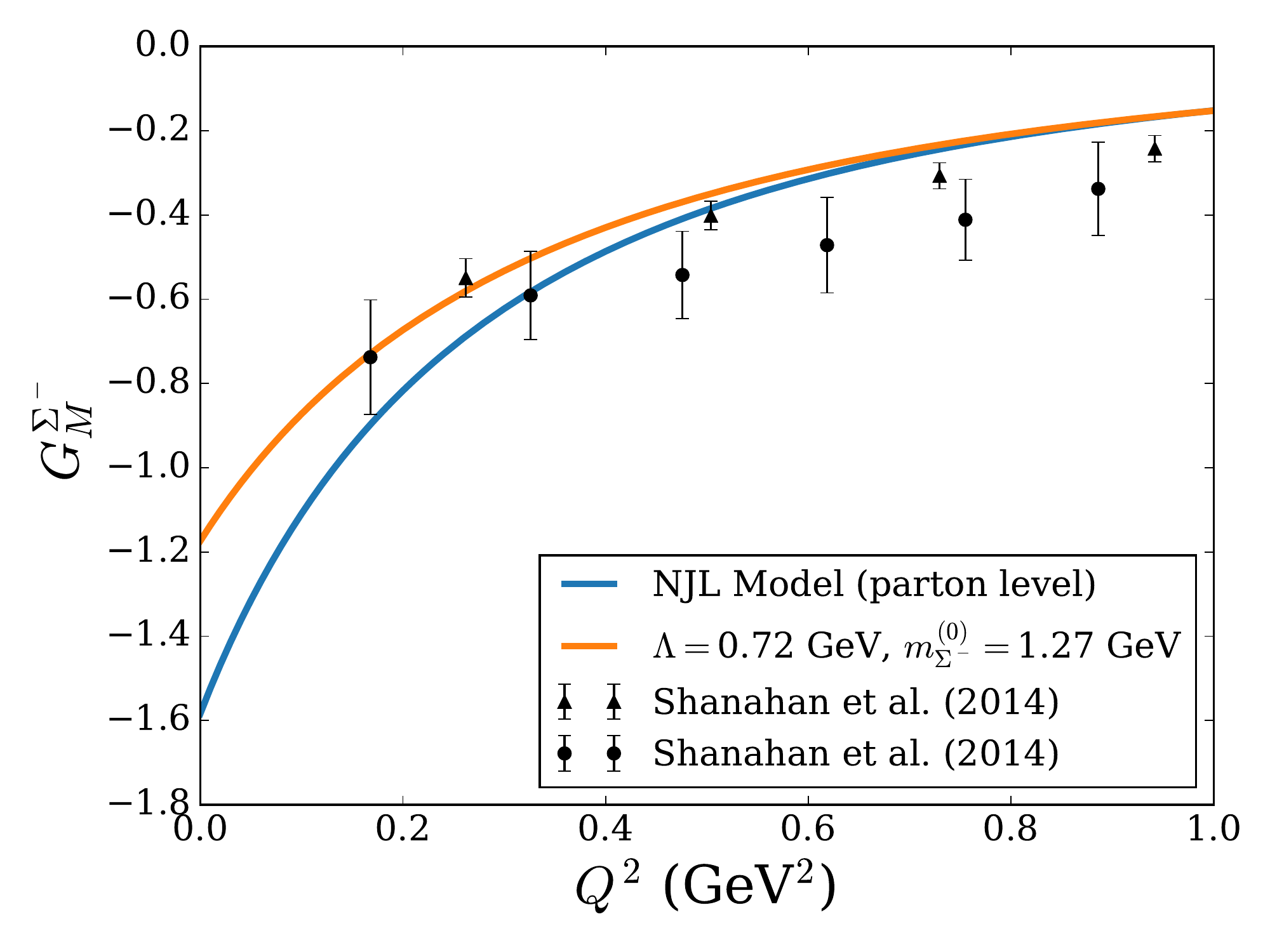}
\end{subfigure}
\begin{subfigure}{0.49\textwidth}
\centering
\includegraphics[scale=0.35]{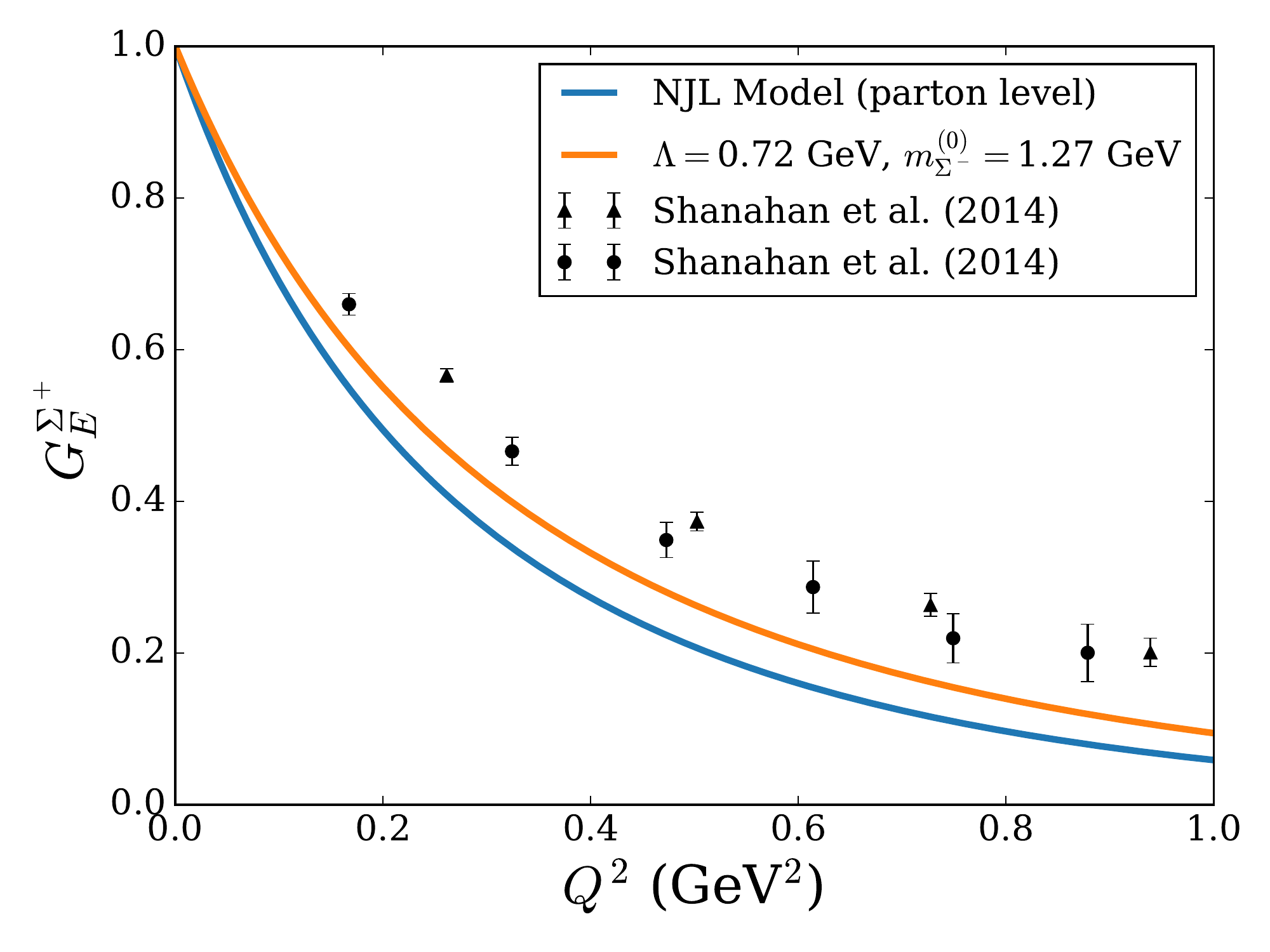}
\end{subfigure}
\begin{subfigure}{0.49\textwidth}
\centering
\includegraphics[scale=0.35]{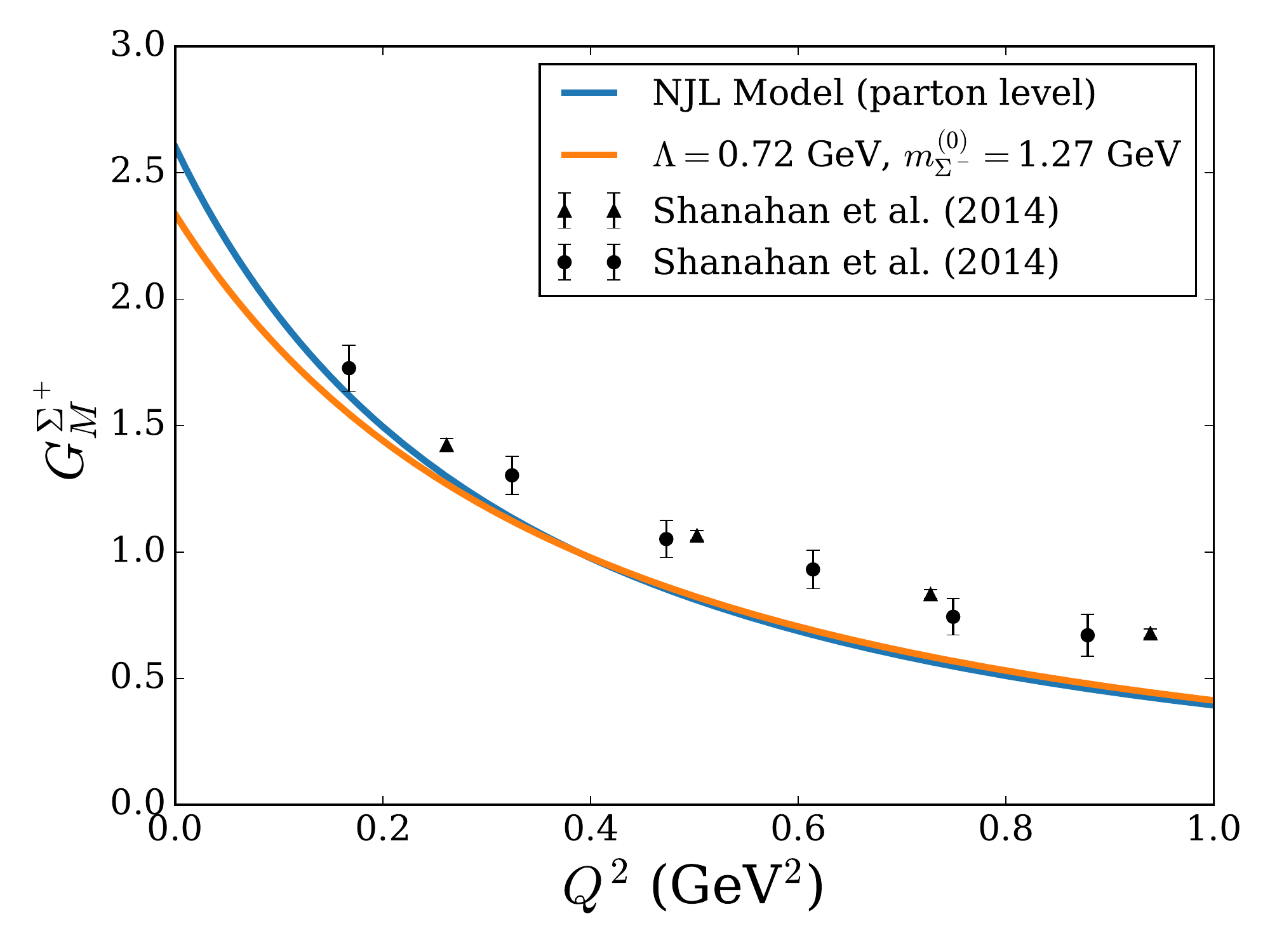}
\end{subfigure}
\vspace{-3mm}
\caption{(Colour online) Electromagnetic form factors for the $\Sigma^-$ 
and $\Sigma^+$. Points with error bars correspond to lattice 
results from~\cite{Shanahan2014a,Shanahan2014b}. Note that the previous 
NJL calculation of the $\Sigma^-$ magnetic moment gives a value of -1.58 $\mu_N$, 
while the method employed in this paper yields a value of 
-1.17 $\mu_N$. The experimental value is quoted as $-1.16 \mu_N$.}\label{fig:4}
\end{figure*}

The bare NJL Model used in the self-consistent evaluation of chiral corrections 
was calculated using the parameters in Table~\ref{tab:1}. 
This set was obtained by fitting the predicted baryon masses for 
the nucleon and $\Xi$ to their experimental values. 
The resulting octet masses are shown in  
Table~\ref{tab:3}, 
Although the predicted values of the $\Lambda$ and $\Sigma$ masses 
differ slightly from the experimental values, as a result of an underestimate 
of the spin-spin interaction in the NJL model,  
the hierarchy of states is correct, that is, $m_N<m_\Lambda<m_\Sigma<m_\Xi$.
\begin{table}[t]
\centering
\begin{tabular}{ c c c c c c c c }
\hline\hline
$\Lambda_{IR}$ &$\Lambda_{UV}$ &$m_l$ &$m_s$ &$G_\pi$ &$G_\rho$ &$G_s$ &$G_a$ \\ \hline
0.24 &0.67 &0.35 &0.52 &14.53 &8.12 &4.11 &3.14 \\ \hline
\end{tabular}
\caption{Chosen NJL model parameters, where all masses and regularization 
parameters are given in units of GeV, and the Lagrangian couplings 
in units of GeV$^{-2}$.}
\label{tab:1}
\end{table}
\begin{table}[h]
\centering
\begin{tabular*}{\columnwidth}{@{\extracolsep{\fill}} c c c c c }
\hline\hline
&$m_N$ &$m_\Lambda$ &$m_\Sigma$ &$m_\Xi$ \\ \hline
NJL$-\Sigma$ &0.940 &1.176 &1.217 &1.318 \\
Experiment &0.940 & 1.116 & 1.193 & 1.318 \\ \hline
\end{tabular*}
\caption{Calculated baryon octet masses (after including the hadron's self energy), 
compared with the experimental values (all in units of GeV).}
\label{tab:3}
\end{table}

This choice of parameters leads to a relatively good agreement between 
the predicted nucleon electromagnetic form factors and the empirical 
parameterization of Kelly~\cite{Kelly2004}, shown in Fig.~\ref{fig:3}. In particular, it  
is certainly of comparable quality to the previous NJL calculation, 
where the pion corrections were 
calculated at the parton level (coloured blue in plots).

As the pion contributes most at low $Q^2$, one may gauge the goodness of fit 
by comparing the predicted low energy observables with their experimental 
values. Table~\ref{tab:5} shows the predicted electric charge radii of 
the studied baryons in comparison to both the previous NJL calculation
(including pion loops on the valence quarks) 
and their respective experimental values. The present calculation
produces an overall description of the nucleon and $\Sigma^-$ charge radii 
of a similar quality to those generated in the earlier, 
parton level model. For the $\Sigma^+$ we stress that the lattice QCD 
result for the charge radius still suffers some systematic uncertainties, as
discussed in the original work.
\begin{table}
\centering
\begin{tabular*}{\columnwidth}{@{\extracolsep{\fill}} c c c c c }
\hline\hline
 \multicolumn{5}{ c }{$\expval{r^2}^\half$} \\ 
 &  $p$ & $n$ & $\Sigma^-$ & $\Sigma^+$ \\ 
\hline 
Prev. NJL Calc. &0.87 &0.38 &0.86 &0.97 \\ 
LFCBM &0.89 &0.41 &0.78 &0.88 \\ 
\hline 
Exp. & 0.84~\cite{Pohl:2010zza} & 0.335 & 0.780 & 0.61(8)~\cite{Shanahan2014b} \\ 
\hline\end{tabular*}
\caption{Comparison of the predicted electric charge radii to experimental 
results for the proton, neutron, $\Sigma^-$ and $\Sigma^+$ baryons. 
Experimental results are taken from~\cite{Kelly2004,Olive2014,Pohl:2010zza}, except for 
the $\Sigma^+$ charge radius, for which there is currently no experimental value. 
In this case, a recent lattice QCD result~\cite{Shanahan2014b} is given instead.
Charge radii are quoted in femtometres.}\label{tab:5}
\end{table}

Comparing the predicted and experimental magnetic moments in Table~\ref{tab:6}, 
we see that the proton magnetic moment agrees with the experimental value, 
while the predicted neutron magnetic magnetic moment is slightly worse than 
the previous NJL Model prediction. The $\Sigma^+$ magnetic moment shows a 
comparable level of agreement with experiment as the previous NJL model.
However, it is for the $\Sigma^-$ magnetic moment that one finds a remarkable
difference when the chiral corrections are implemented correctly. Whereas 
evaluating the loops on individual quarks leads to $\mu(\Sigma^-) = -1.58 \mu_N$, 
far larger than the empirical value($-1.160 (25) \mu_N$), when implemented 
correctly at the hadronic level one finds $\mu(\Sigma^-) = - 1.17 \mu_N$, 
which is in excellent agreement with experiment.
The reason for the overestimate in the $\Sigma^-$ case is that the pion cloud 
on a $d$-quark dramatically increases its magnetic moment. 
The relative increase for a $u$-quark is also significant and all three quarks give
a negative correction to the $\Sigma^-$ as the valence $u$-quark has 
spin down.

Finally, we observe that for the $\Sigma$ hyperon form factors 
up to 1 GeV$^2$ (shown in Fig.~\ref{fig:4}),
the level of agreement 
between the calculations including pionic corrections 
at the hadronic level and recent 
lattice results including chiral corrections are 
typically as good as, or better than the calculations made at the quark level.
\begin{table}
\centering
\begin{tabular*}{\columnwidth}{@{\extracolsep{\fill}} c c c c c }
\hline\hline
 \multicolumn{5}{ c }{$\mu$} \\ 
 & $p$ & $n$ & $\Sigma^-$ & $\Sigma^+$ \\ 
\hline 
Prev. NJL Calc. &2.78 &-1.81 &-1.58 &2.60 \\ 
LFCBM &2.78 &-1.71 &-1.17 &2.33 \\ 
\hline 
Exp. & 2.793 & -1.913 & -1.160(25) & 2.458(10)\\ 
\hline\end{tabular*}
\caption{Comparison of the predicted magnetic moments to experimental 
results for the proton, neutron, $\Sigma^-$ and $\Sigma^+$ baryons. 
Experimental results are taken from~\cite{Kelly2004,Olive2014}. 
Magnetic moments are in units of nuclear magnetons 
($\mu_N=e/2m_N$).}\label{tab:6}
\end{table}
%

\section{Conclusion}\label{sec:5}
In this work we have investigated the practical importance of a 
correct implementation of chiral symmetry for the electromagnetic 
form factors of the nucleons and  
$\Sigma$ hyperons. The NJL model was used to evaluate the 
form factors of the underlying quark structure, while the pion 
loop corrections were evaluated at the baryon level using the 
light-front cloudy bag model. The results were compared with
experimental data where available, or with the results
of a recent lattice QCD simulation to which chiral corrections 
had been applied if no experimental value existed. 

Remarkably for the proton and neutron, there was little 
practical difference between the results of 
implementing the chiral corrections at the hadron or parton level. 
On the other hand, for the magnetic form factor of the $\Sigma^-$, there was a 
dramatic improvement when the pion corrections were evaluated correctly. 
In particular, the $\Sigma^-$ magnetic moment is reduced by roughly 30\% 
compared with an evaluation at the individual quark level, with the new result 
now in excellent agreement with experiment.

In summary, while it is convenient to evaluate pion loop corrections on 
individual quarks, independent of the hadronic environment, the results 
reported here illustrate very clearly that such an 
approach can deliver (at least for the hyperons studied)
inaccurate results. Worse, there seems to be no obvious way to predict 
ahead of time whether or not the calculated values may be expected to be 
reliable. Hence, it is clear that a theoretically consistent, 
reliable result may only be obtained by performing the 
chiral corrections at the hadron level, as is done here. 

\section*{Acknowledgements}
We would like to thank W. Bentz for helpful comments on the manuscript.
This work was supported by the Australian Research Council through 
grant DP150103101 (AWT) as well as through the ARC Centre of Excellence for 
Particle Physics at the Terascale, CE110001104.
\section*{References}
\bibliography{bibliography}

\begin{thebibliography}{10}
\expandafter\ifx\csname url\endcsname\relax
  \def\url#1{\texttt{#1}}\fi
\expandafter\ifx\csname urlprefix\endcsname\relax\def\urlprefix{URL }\fi
\expandafter\ifx\csname href\endcsname\relax
  \def\href#1#2{#2} \def\path#1{#1}\fi

\bibitem{Isgur:1987ht}
N.~Isgur, G.~Karl, J.~Soffer, {Zeros in the Nucleon Form-factors and the Quark
  Model}, Phys. Rev. D35 (1987) 1665--1667.
\newblock \href {http://dx.doi.org/10.1103/PhysRevD.35.1665}
  {\path{doi:10.1103/PhysRevD.35.1665}}.

\bibitem{Melde:2007zz}
T.~Melde, K.~Berger, L.~Canton, W.~Plessas, R.~F. Wagenbrunn, {Electromagnetic
  nucleon form factors in instant and point form}, Phys. Rev. D76 (2007)
  074020.
\newblock \href {http://dx.doi.org/10.1103/PhysRevD.76.074020}
  {\path{doi:10.1103/PhysRevD.76.074020}}.

\bibitem{Chodos:1974pn}
A.~Chodos, R.~L. Jaffe, K.~Johnson, C.~B. Thorn, {Baryon Structure in the Bag
  Theory}, Phys. Rev. D10 (1974) 2599.
\newblock \href {http://dx.doi.org/10.1103/PhysRevD.10.2599}
  {\path{doi:10.1103/PhysRevD.10.2599}}.

\bibitem{Cloet:2013gva}
I.~C. Cloet, C.~D. Roberts, A.~W. Thomas, {Revealing dressed-quarks via the
  proton's charge distribution}, Phys. Rev. Lett. 111 (2013) 101803.
\newblock \href {http://arxiv.org/abs/1304.0855} {\path{arXiv:1304.0855}},
  \href {http://dx.doi.org/10.1103/PhysRevLett.111.101803}
  {\path{doi:10.1103/PhysRevLett.111.101803}}.

\bibitem{Ishii:1993np}
N.~Ishii, W.~Bentz, K.~Yazaki, {Faddeev approach to the nucleon in the
  Nambu-Jona-Lasinio (NJL) model}, Phys. Lett. B301 (1993) 165--169.
\newblock \href {http://dx.doi.org/10.1016/0370-2693(93)90683-9}
  {\path{doi:10.1016/0370-2693(93)90683-9}}.

\bibitem{Schroeder:1999fr}
O.~Schroeder, H.~Reinhardt, H.~Weigel, {Nucleon structure functions in the
  three flavor NJL soliton model}, Nucl. Phys. A651 (1999) 174.
\newblock \href {http://arxiv.org/abs/hep-ph/9902322}
  {\path{arXiv:hep-ph/9902322}}, \href
  {http://dx.doi.org/10.1016/S0375-9474(99)00131-1}
  {\path{doi:10.1016/S0375-9474(99)00131-1}}.

\bibitem{Cloet:2014rja}
I.~C. Cloët, W.~Bentz, A.~W. Thomas, {Role of diquark correlations and the
  pion cloud in nucleon elastic form factors}, Phys. Rev. C90 (2014) 045202.
\newblock \href {http://arxiv.org/abs/1405.5542} {\path{arXiv:1405.5542}},
  \href {http://dx.doi.org/10.1103/PhysRevC.90.045202}
  {\path{doi:10.1103/PhysRevC.90.045202}}.

\bibitem{Carrillo-Serrano:2016igi}
M.~E. Carrillo-Serrano, W.~Bentz, I.~C. Cloët, A.~W. Thomas, {Baryon Octet
  Electromagnetic Form Factors in a confining NJL model}, Phys. Lett. B759
  (2016) 178--183.
\newblock \href {http://arxiv.org/abs/1603.02741} {\path{arXiv:1603.02741}},
  \href {http://dx.doi.org/10.1016/j.physletb.2016.05.065}
  {\path{doi:10.1016/j.physletb.2016.05.065}}.

\bibitem{Ebert:1996vx}
D.~Ebert, T.~Feldmann, H.~Reinhardt, {Extended NJL model for light and heavy
  mesons without q - anti-q thresholds}, Phys. Lett. B388 (1996) 154--160.
\newblock \href {http://arxiv.org/abs/hep-ph/9608223}
  {\path{arXiv:hep-ph/9608223}}, \href
  {http://dx.doi.org/10.1016/0370-2693(96)01158-6}
  {\path{doi:10.1016/0370-2693(96)01158-6}}.

\bibitem{Weinberg:1966kf}
S.~Weinberg, {Pion scattering lengths}, Phys. Rev. Lett. 17 (1966) 616--621.
\newblock \href {http://dx.doi.org/10.1103/PhysRevLett.17.616}
  {\path{doi:10.1103/PhysRevLett.17.616}}.

\bibitem{Pagels:1974se}
H.~Pagels, {Departures from Chiral Symmetry: A Review}, Phys. Rept. 16 (1975)
  219.
\newblock \href {http://dx.doi.org/10.1016/0370-1573(75)90039-3}
  {\path{doi:10.1016/0370-1573(75)90039-3}}.

\bibitem{Thomas1999}
A.~Thomas, G.~Krein,
  \href{http://dx.doi.org/10.1016/S0370-2693(99)00455-4}{Chiral corrections in
  hadron spectroscopy}, Physics Letters B 456~(1) (1999) 5--8.
\newblock \href {http://dx.doi.org/10.1016/s0370-2693(99)00455-4}
  {\path{doi:10.1016/s0370-2693(99)00455-4}}.
\newline\urlprefix\url{http://dx.doi.org/10.1016/S0370-2693(99)00455-4}

\bibitem{Thomas2000}
A.~Thomas, G.~Krein,
  \href{http://dx.doi.org/10.1016/S0370-2693(00)00426-3}{Chiral aspects of
  hadron structure}, Physics Letters B 481~(1) (2000) 21--25.
\newblock \href {http://dx.doi.org/10.1016/s0370-2693(00)00426-3}
  {\path{doi:10.1016/s0370-2693(00)00426-3}}.
\newline\urlprefix\url{http://dx.doi.org/10.1016/S0370-2693(00)00426-3}

\bibitem{Manohar:1983md}
A.~Manohar, H.~Georgi, {Chiral Quarks and the Nonrelativistic Quark Model},
  Nucl. Phys. B234 (1984) 189--212.
\newblock \href {http://dx.doi.org/10.1016/0550-3213(84)90231-1}
  {\path{doi:10.1016/0550-3213(84)90231-1}}.

\bibitem{CarrilloSerrano2014}
M.~E. Carrillo-Serrano, I.~C. Clo\"et, A.~W. Thomas,
  \href{https://doi.org/10.1103%2Fphysrevc.90.064316}{{SU}(3)-flavor breaking
  in octet baryon masses and axial couplings}, Physical Review C 90~(6).
\newblock \href {http://dx.doi.org/10.1103/physrevc.90.064316}
  {\path{doi:10.1103/physrevc.90.064316}}.
\newline\urlprefix\url{https://doi.org/10.1103%2Fphysrevc.90.064316}

\bibitem{Cloet2014}
I.~C. Clo\"et, W.~Bentz, A.~W. Thomas,
  \href{http://dx.doi.org/10.1103/PhysRevC.90.045202}{Role of diquark
  correlations and the pion cloud in nucleon elastic form factors}, Physical
  Review C 90~(4).
\newblock \href {http://dx.doi.org/10.1103/physrevc.90.045202}
  {\path{doi:10.1103/physrevc.90.045202}}.
\newline\urlprefix\url{http://dx.doi.org/10.1103/PhysRevC.90.045202}

\bibitem{CarrilloSerrano2016}
M.~E. Carrillo-Serrano, W.~Bentz, I.~C. Clo\"et, A.~W. Thomas,
  \href{http://dx.doi.org/10.1016/j.physletb.2016.05.065}{Baryon octet
  electromagnetic form factors in a confining {NJL} model}, Physics Letters B
  759 (2016) 178--183.
\newblock \href {http://dx.doi.org/10.1016/j.physletb.2016.05.065}
  {\path{doi:10.1016/j.physletb.2016.05.065}}.
\newline\urlprefix\url{http://dx.doi.org/10.1016/j.physletb.2016.05.065}

\bibitem{Nambu:1961tp}
Y.~Nambu, G.~Jona-Lasinio, {Dynamical Model of Elementary Particles Based on an
  Analogy with Superconductivity. 1.}, Phys. Rev. 122 (1961) 345--358.
\newblock \href {http://dx.doi.org/10.1103/PhysRev.122.345}
  {\path{doi:10.1103/PhysRev.122.345}}.

\bibitem{Nambu:1961fr}
Y.~Nambu, G.~Jona-Lasinio, {DYNAMICAL MODEL OF ELEMENTARY PARTICLES BASED ON AN
  ANALOGY WITH SUPERCONDUCTIVITY. II}, Phys. Rev. 124 (1961) 246--254.
\newblock \href {http://dx.doi.org/10.1103/PhysRev.124.246}
  {\path{doi:10.1103/PhysRev.124.246}}.

\bibitem{Miller2002}
G.~A. Miller, \href{http://dx.doi.org/10.1103/PhysRevC.66.032201}{Light front
  cloudy bag model: Nucleon electromagnetic form factors}, Physical Review C
  66~(3).
\newblock \href {http://dx.doi.org/10.1103/physrevc.66.032201}
  {\path{doi:10.1103/physrevc.66.032201}}.
\newline\urlprefix\url{http://dx.doi.org/10.1103/PhysRevC.66.032201}

\bibitem{Theberge1980}
S.~Th{\'{e}}berge, A.~W. Thomas, G.~A. Miller,
  \href{https://doi.org/10.1103%2Fphysrevd.22.2838}{Pionic corrections to the
  {MIT} bag model: The (3, 3) resonance}, Physical Review D 22~(11) (1980)
  2838--2852.
\newblock \href {http://dx.doi.org/10.1103/physrevd.22.2838}
  {\path{doi:10.1103/physrevd.22.2838}}.
\newline\urlprefix\url{https://doi.org/10.1103%2Fphysrevd.22.2838}

\bibitem{Miller1980}
G.~Miller, A.~Thomas, S.~Th{\'{e}}berge,
  \href{https://doi.org/10.1016%2F0370-2693%2880%2990428-1}{Pion-nucleon
  scattering in the brown-rho bag model}, Physics Letters B 91~(2) (1980)
  192--195.
\newblock \href {http://dx.doi.org/10.1016/0370-2693(80)90428-1}
  {\path{doi:10.1016/0370-2693(80)90428-1}}.
\newline\urlprefix\url{https://doi.org/10.1016%2F0370-2693%2880%2990428-1}

\bibitem{Miller:1984em}
G.~A. Miller, {Building the Nucleus From Quarks: the Cloudy bag Model and the
  Quark Description of the Nucleon-nucleon Wave Functions}, Int. Rev. Nucl.
  Phys. 1 (1984) 189--323.
\newblock \href {http://dx.doi.org/10.1142/9789814415132_0003}
  {\path{doi:10.1142/9789814415132_0003}}.

\bibitem{Thomas1984}
A.~W. Thomas, \href{https://doi.org/10.1007%2F978-1-4613-9892-9_1}{Chiral
  symmetry and the {BAG} model: A new starting point for nuclear physics}, in:
  Advances in Nuclear Physics, Springer Nature, 1984, pp. 1--137.
\newblock \href {http://dx.doi.org/10.1007/978-1-4613-9892-9_1}
  {\path{doi:10.1007/978-1-4613-9892-9_1}}.
\newline\urlprefix\url{https://doi.org/10.1007%2F978-1-4613-9892-9_1}

\bibitem{Schlumpf92}
F.~Schlumpf, {Relativistic constituent quark model for baryons, Doctoral
  Thesis}.

\bibitem{Matevosyan2005}
H.~H. Matevosyan, G.~A. Miller, A.~W. Thomas,
  \href{http://dx.doi.org/10.1103/PhysRevC.71.055204}{Comparison of nucleon
  form factors from lattice {QCD} against the light front cloudy bag model and
  extrapolation to the physical mass regime}, Physical Review C 71~(5).
\newblock \href {http://dx.doi.org/10.1103/physrevc.71.055204}
  {\path{doi:10.1103/physrevc.71.055204}}.
\newline\urlprefix\url{http://dx.doi.org/10.1103/PhysRevC.71.055204}

\bibitem{McKenney:2015xis}
J.~R. McKenney, N.~Sato, W.~Melnitchouk, C.-R. Ji, {Pion structure function
  from leading neutron electroproduction and SU(2) flavor asymmetry}, Phys.
  Rev. D93~(5) (2016) 054011.
\newblock \href {http://arxiv.org/abs/1512.04459} {\path{arXiv:1512.04459}},
  \href {http://dx.doi.org/10.1103/PhysRevD.93.054011}
  {\path{doi:10.1103/PhysRevD.93.054011}}.

\bibitem{Thomas:1983fh}
A.~W. Thomas, {A Limit on the Pionic Component of the Nucleon Through SU(3)
  Flavor Breaking in the Sea}, Phys. Lett. B126 (1983) 97--100.
\newblock \href {http://dx.doi.org/10.1016/0370-2693(83)90026-6}
  {\path{doi:10.1016/0370-2693(83)90026-6}}.

\bibitem{Melnitchouk:1998rv}
W.~Melnitchouk, J.~Speth, A.~W. Thomas, {Dynamics of light anti-quarks in the
  proton}, Phys. Rev. D59 (1998) 014033.
\newblock \href {http://arxiv.org/abs/hep-ph/9806255}
  {\path{arXiv:hep-ph/9806255}}, \href
  {http://dx.doi.org/10.1103/PhysRevD.59.014033}
  {\path{doi:10.1103/PhysRevD.59.014033}}.

\bibitem{deSwart1963}
J.~J. de~Swart, \href{http://dx.doi.org/10.1103/RevModPhys.35.916}{The octet
  model and its clebsch-gordan coefficients}, Reviews of Modern Physics 35~(4)
  (1963) 916--939.
\newblock \href {http://dx.doi.org/10.1103/revmodphys.35.916}
  {\path{doi:10.1103/revmodphys.35.916}}.
\newline\urlprefix\url{http://dx.doi.org/10.1103/RevModPhys.35.916}

\bibitem{Theberge1982}
S.~Th\'eberge, The cloudy bag model, Ph.D. thesis (March 1982).

\bibitem{Hecht:2002ej}
M.~B. Hecht, M.~Oettel, C.~D. Roberts, S.~M. Schmidt, P.~C. Tandy, A.~W.
  Thomas, {Nucleon mass and pion loops}, Phys. Rev. C65 (2002) 055204.
\newblock \href {http://arxiv.org/abs/nucl-th/0201084}
  {\path{arXiv:nucl-th/0201084}}, \href
  {http://dx.doi.org/10.1103/PhysRevC.65.055204}
  {\path{doi:10.1103/PhysRevC.65.055204}}.

\bibitem{Young:2002cj}
R.~D. Young, D.~B. Leinweber, A.~W. Thomas, S.~V. Wright, {Chiral analysis of
  quenched baryon masses}, Phys. Rev. D66 (2002) 094507.
\newblock \href {http://arxiv.org/abs/hep-lat/0205017}
  {\path{arXiv:hep-lat/0205017}}, \href
  {http://dx.doi.org/10.1103/PhysRevD.66.094507}
  {\path{doi:10.1103/PhysRevD.66.094507}}.

\bibitem{Kelly2004}
J.~J. Kelly, \href{https://doi.org/10.1103%2Fphysrevc.70.068202}{Simple
  parametrization of nucleon form factors}, Physical Review C 70~(6).
\newblock \href {http://dx.doi.org/10.1103/physrevc.70.068202}
  {\path{doi:10.1103/physrevc.70.068202}}.
\newline\urlprefix\url{https://doi.org/10.1103%2Fphysrevc.70.068202}

\bibitem{Shanahan2014a}
P.~Shanahan, R.~Horsley, Y.~Nakamura, D.~Pleiter, P.~Rakow, G.~Schierholz,
  H.~St\"{u}ben, A.~Thomas, R.~Young, J.~Z. and,
  \href{https://doi.org/10.1103%2Fphysrevd.89.074511}{Magnetic form factors of
  the octet baryons from lattice {QCD} and chiral extrapolation}, Physical
  Review D 89~(7).
\newblock \href {http://dx.doi.org/10.1103/physrevd.89.074511}
  {\path{doi:10.1103/physrevd.89.074511}}.
\newline\urlprefix\url{https://doi.org/10.1103%2Fphysrevd.89.074511}

\bibitem{Shanahan2014b}
P.~Shanahan, R.~Horsley, Y.~Nakamura, D.~Pleiter, P.~Rakow, G.~Schierholz,
  H.~St\"{u}ben, A.~Thomas, R.~Young, J.~Z. and,
  \href{https://doi.org/10.1103%2Fphysrevd.90.034502}{Electric form factors of
  the octet baryons from lattice {QCD} and chiral extrapolation}, Physical
  Review D 90~(3).
\newblock \href {http://dx.doi.org/10.1103/physrevd.90.034502}
  {\path{doi:10.1103/physrevd.90.034502}}.
\newline\urlprefix\url{https://doi.org/10.1103%2Fphysrevd.90.034502}

\bibitem{Pohl:2010zza}
R.~Pohl, et~al., {The size of the proton}, Nature 466 (2010) 213--216.
\newblock \href {http://dx.doi.org/10.1038/nature09250}
  {\path{doi:10.1038/nature09250}}.

\bibitem{Olive2014}
K.~Olive, \href{https://doi.org/10.1088%2F1674-1137%2F38%2F9%2F090001}{Review
  of particle physics}, Chinese Physics C 38~(9) (2014) 090001.
\newblock \href {http://dx.doi.org/10.1088/1674-1137/38/9/090001}
  {\path{doi:10.1088/1674-1137/38/9/090001}}.
\newline\urlprefix\url{https://doi.org/10.1088%2F1674-1137%2F38%2F9%2F090001}

\end{thebibliography}

\end{fmffile}
\end{document}